\documentclass[fleqn,usenatbib,useAMS,onecolumn]{mnras}
\usepackage{graphicx}	
\usepackage{amsmath}	
\usepackage{amssymb}	
\usepackage{multicol}        
\usepackage{bm}		
\usepackage{pdflscape}
\usepackage[T1]{fontenc}
\usepackage{ae,aecompl}
\usepackage{newtxtext,newtxmath}
\usepackage{calc}
\usepackage{pifont}
\DeclareRobustCommand{\VAN}[3]{#2}
\let\VANthebibliography\thebibliography
\def\thebibliography{\DeclareRobustCommand{\VAN}[3]{##3}\VANthebibliography}

\usepackage{graphicx}	
\usepackage{amsmath}	

\def \aj {{\rm {AJ}}}

\def \apj {{\rm {ApJ}}}
\def \icarus {{\rm {Icarus}}}
\def \apjs {{\rm {ApJS}}}

\def \aap {{\rm {A\&A}}}

\def \mnras {{\rm {MNRAS}}}

\def \apjl{\rm {ApJL}}
\def \nat{\rm {Nat.}}

\def \deg{$^\circ$}
\def \mj{\,$M_{\rm Jup}$\,}

\def \micron{\,$\mu$m\,}

\def \msun{\,$M_\odot$\,}

\newcommand{\ms}{\mbox{\,m s$^{-1}$}}
\newcommand{\msyr}{\mbox{\,m s$^{-1}$yr$^{-1}$}}

\newcommand{\ind}{$\varepsilon$ Ind A }
\newcommand{\eri}{$\varepsilon$ Eridani }
\newcommand{\as}{$''$}

\title[Revised orbits of the two nearest Jupiters]{Revised orbits of the two nearest Jupiters}

\author[F. Feng et al.]{
Fabo Feng,$^{1,2}$\thanks{E-mail: ffeng@sjtu.edu.cn (TDLI)}
R. Paul Butler,$^{3}$
Steven S. Vogt,$^{4}$
Bradford Holden$^{4}$
\newauthor
and Yicheng Rui$^{1}$
\\
$^{1}$Tsung-Dao Lee Institute, Shanghai Jiao Tong University, Shengrong Road 520, Shanghai, 201210, People's Republic Of China\\
$^{2}$School of Physics and Astronomy, Shanghai Jiao Tong University, 800 Dongchuan Road, Shanghai 200240, People's Republic of China\\
$^{3}$Earth and Planets Laboratory, Carnegie Institution for Science, Washington, DC 20015, USA\\
$^{4}$UCO/Lick Observatory, University of California, Santa Cruz, CA 95064,USA
}

\date{Accepted XXX. Received YYY; in original form ZZZ}

\pubyear{2023}

\begin{document}
\label{firstpage}
\pagerange{\pageref{firstpage}--\pageref{lastpage}}
\maketitle

\begin{abstract}
With its near-to-mid-infrared high contrast imaging capabilities,
  JWST is ushering us into a golden age of directly imaging
  Jupiter-like planets. As the two closest cold Jupiters, \ind b and
  \eri b have sufficiently wide orbits and adequate infrared emissions to be detected by JWST. To detect more Jupiter-like planets for direct imaging, we develop
a GOST-based method to analyze radial velocity data and
multiple Gaia data releases simultaneously. Without approximating instantaneous
astrometry by catalog astrometry, this approach enables the use of
multiple Gaia data releases for detection of both short-period and
long-period planets. We determine a mass of $2.96_{-0.38}^{+0.41}$\mj and a period of
  $42.92_{-4.09}^{+6.38}$\,yr for \ind b. We also find a mass of $0.76_{-0.11}^{+0.14}$\mj, a period of $7.36_{-0.05}^{+0.04}$\,yr,
and an eccentricity of 0.26$_{-0.04}^{+0.04}$ for \eri b. The
eccentricity differs from that given by some previous solutions probably due
to the sensitivity of orbital eccentricity to noise modeling. Our work
refines the constraints on orbits and masses of the two nearest
Jupiters and demonstrate the feasibility of using multiple Gaia
  data releases to constrain Jupiter-like planets.
\end{abstract}




\section{Introduction}\label{sec:intro}
The detection and characterization of Jupiter-like planets in
extrasolar systems are crucial for our understanding of their formation and
evolution as well as their role in shaping the system architecture and
habitability \citep{stevenson88,lunine01,tsiganis05,horner20}. While
the current ground-based facilities are mainly sensitive to young
Jupiters, the Mid-Infrared Instrument (MIRI; \citealt{rieke15}) mounted on the James Webb Space
Telescope (JWST) is optimal for imaging cold Jupiters, which are far
more abundant than young Jupiters. Equipped with a
coronagraph, MIRI has an imaging observing mode from 4.7 to
27.9\micron, sensitive to the mid-infrared emission from cold Jupiters
with equilibrium temperature of less than 200\,K \citep{bouchet15}. On the other hand,
the JWST Near Infrared Spectrograph (NIRSpec; \citealt{jakobsen22}) conducts
medium-resolution spectroscopic observation from 0.6 to
5.3\micron. The differential Doppler shift of planet relative to its
host star could be used to disentangle planetary and stellar spectrums for planet imaging \citep{llop21}.

Considering that high contrast imaging typically requires large
planet-star separation and high planetary mass, nearby cold Jupiters
are optimal targets for JWST imaging. However, the orbital periods of
cold Jupiters are as long as decades, beyond most of the individual
radial velocity (RV), astrometric, and transit surveys. Hence, it is
important to form extremely long observational time span (or baseline)
by combining different types of data. One successful application of
such synergistic approach is combined analyses of legacy RV data and the
astrometric data from Hipparcos \citep{perryman97,leeuwen07} and Gaia
data releases \citep{gaia16,gaia18,gaia20,gaia22}. This approach not only provides decade-long baseline to constrain the orbit of a cold
Jupiter but also fully constrain planetary mass and orbit due to the
complementarity between RV and astrometry. Assuming that the catalog
astrometry approximates the instantaneous astrometry at the reference
epoch, one can use the difference between Hipparcos and Gaia
astrometry to constrain the nonlinear motion of a star (or reflex motion) induced by its
companion \citep{snellen18,brandt18,feng19b,kervella19}. However,
under this assumption, Gaia DR2 and DR3 are not independent and thus
cannot be used simultaneously to constrain planetary orbit. Moreover, this
assumption is not appropriate for constraining orbits with periods
comparable or shorter than the observational time span of the epoch
data used to produce an astrometric catalog.

To use multiple Gaia data releases to constrain both short and long-period planets, we simulate the Gaia epoch data using the Gaia Observation Forecast Tool
(GOST\footnote{GOST website: \url{https://gaia.esac.esa.int/gost/} and
 GOST user manual: \url{https://gaia.esac.esa.int/gost/docs/gost_software_user_manual.pdf}})
and fit the synthetic data by a linear astrometric model, corresponding to the Gaia five-parameter solution. The difference
between the fitted and the catalog astrometry constrains the
nonlinear reflex motion induced by planets. In this work, we apply this
approach to constrain the orbits and masses of the two nearest cold
Jupiters, \ind b and \eri b. 

This paper is structured as follows. In section \ref{sec:data}, we
introduce the RV and astrometry data measured for \ind and \eri. We
then describe the modeling and statistical methods used in our
analyses in section \ref{sec:method}. We give the orbital solutions
for the two planets and compare our solutions with previous ones in
section \ref{sec:results}. Finally, we discuss and conclude in section \ref{sec:conclusion}.

\section{RV and astrometry data}\label{sec:data}
\ind is a K-type star with a mass of 0.754$\pm$0.038\msun
\citep{demory09}, located at a heliocentric distance of 3.7\,pc. It
hosts two brown dwarfs \citep{smith03} and a Jupiter-like planet \citep{feng19b}. It has been observed by high precision spectrographs including
the High Accuracy Radial Velocity Planet Searcher (HARPS;
\citealt{pepe00}) mounted on the ESO La Silla 3.6m telescope, the ESO UV-visual echelle spectrograph (UVES)
on Unit 2 of the Very Large Telescope (VLT) array \citep{dekker00},
and the Coud\'e Echelle Spectrograph (CES) at the 1.4 m telescope in
La Silla, Chile. We use ``HARPSpre'' and ``HARPSpost'' to denote
HARPS data obtained before and after the fibre change in 2015
\citep{curto15}. The HARPS data are reduced by \cite{trifonov20} using the SERVAL pipeline \citep{zechmeister18}. The RV data obtained
by the Long Camera (LC) and the Very Long Camera (VLC) were released
by \cite{zechmeister13}.

As a K-type star with a mass of 0.82$\pm$0.02\msun\citep{gonzalez10},
\eri is the third closest star system to the Earth. It is likely that
a Jupiter analog orbits around this star on a wide orbit
\citep{hatzes00,butler06,mawet19}.  It has been observed by the CES LC
and VLC \citep{zechmeister13}, the Lick Observatory Hamilton echelle
spectrometer \citep{vogt87}, the Automated Planet Finder (APF; \citealt{vogt14}),
the HIRES spectrometer \citep{vogt94} at the Keck observatory, the
HARPS for the Northern hemisphere (HARPS-N or HARPN;
\citealt{cosentino12}) installed at the Italian Telescopio Nazionale Galileo (TNG), the EXtreme PREcision
Spectrograph (EXPRES; \citealt{jurgenson16}) installed at the 4.3 m
Lowell Discovery Telescope (LDT; \citealt{levine12}). By accounting
for the RV offets caused by the updates of Lick Hamilton spectrograph,
we use ``Lick6'', ``Lick8'', and ``Lick13'' to denote multiple data sets,
following the convention given by \cite{fischer13}. We use the APF
and HIRES data reduced by the standard CPS pipeline \citep{howard10}
and released by \cite{mawet19} as well as the other APF data
reduced using the pipeline developed by \cite{butler96} and the HIRES
data released by \cite{butler17}. We use ``APFh'' and
``HIRESh'' to denote the former sets and use ``APFp'' and ``HIRESp''
to denote the latter. The APFp data set is presented in Table
\ref{tab:apfp}. The EXPRES data is from \cite{roettenbacher22}
and the HARPS data is obtained from \cite{trifonov20}.

We obtain the Hipparcos epoch data for \ind and \eri from the new Hipparcos reduction
\citep{leeuwen07}. We use the Gaia second and third data releases (DR2
and DR3; \citealt{gaia20,gaia22}) as well as the epoch data generated by GOST. The Gaia first data release (DR1) is not used because it is derived partly from
the Hipparcos data \citep{michalik15} and is thus not treated as independent of the Hipparcos data. 

\section{Method}\label{sec:method}
Considering that the Gaia intermediate astrometric data (or epoch
data) are not available in the current Gaia data releases, techniques have been developed to use the difference
between Gaia and Hipparcos catalog data to constrain the orbits of substellar companions
\citep{brandt19,kervella19,feng22}. Without using epoch data, previous
studies are limited by approximating the simultaneous astrometry at the reference
epoch with the catalog astrometry \citep{feng22}. In other words, a linear function is not appropriate to model
the center (position) and tangential line (proper motion) of the
orbital arc of the stellar reflex motion induced by short-period companions.

To avoid the above assumptions, we use GOST to predict the Gaia observation epochs for a given star. Considering that the models of RV
and reflex motion are already introduced in the previous papers
authored by some of us \citep{feng19b,feng21}, we introduce the newly developed technique of
using Hipparcos intermediate data and Gaia GOST data as follows.
\begin{itemize}
\item {\bf Obtain data.} For a given target, we obtain the revised Hipparcos intermediate
  data IAD from \cite{leeuwen07} and Gaia GOST data, including the scan angle $\psi$, the
  along-scan (AL) parallax factor $f^{\rm AL}$, and the observation
  time at barycenter. In previous studies, different versions of the Hipparcos catalog data have been recommended \citep{esa97,leeuwen07}. However, in our research, we find that the choice of the Hipparcos version has minimal impact on our orbital solutions. We attribute this to several reasons:
\begin{itemize}
\item Our approach involves modeling the systematics in Hipparcos IAD using offsets and jitters instead of calibrating them a priori, as described in \cite{brandt23}. By incorporating these offsets and jitters, we account for the systematic effects in the data, making our solutions less sensitive to the specific Hipparcos version used.

\item The astrometric precision of Hipparcos is considerably inferior
  to that of Gaia. Additionally, the time difference between Gaia and
  Hipparcos is much greater than the duration covered by Gaia data
  releases (DRs). Consequently, when it comes to constraining long-period orbits, the crucial factor is the temporal baseline between Hipparcos and Gaia, rather than the particular version of the Hipparcos catalog.

\item For short-period orbits, it is the curvature of the Hipparcos
  IAD that primarily constrains the orbit, rather than the absolute
  offset. Hence a calibration of the offset of Hipparcos IAD 
  becomes less critical in determining short-period orbits.
\end{itemize}
\item {\bf Model astrometry of target system barycenter (TSB) at Gaia
    DR3 reference epoch.} We model the astrometry of the TSB at the
  Gaia DR3 epoch J2016.0 ($t_{\rm DR3}$) as follows:
  \begin{eqnarray}
    \alpha^b_{\rm DR3}&=&\alpha_{\rm DR3}-\frac{\Delta\alpha_*}{\cos\delta_{\rm DR3}}~,\\
    \delta^b_{\rm DR3}&=&\delta_{\rm DR3}-\Delta\delta~,\\
    \varpi^b_{\rm DR3}&=&\varpi_{\rm DR3}-\Delta\varpi~,\\
    \mu^b_{\alpha {\rm DR3}}&=&\mu_{\alpha {\rm DR3}}-\Delta\mu_\alpha~,\\
    \mu^b_{\delta {\rm DR3}}&=&\mu_{\delta {\rm DR3}}-\Delta\mu_\delta~,
  \end{eqnarray}
where $\alpha$, $\delta$, $\varpi$,
$\mu_\alpha$, and $\mu_\delta$ are R.A., decl., parallax, and proper
motion in R.A and decl.\footnote{Note that $\mu_\alpha$ is defined as
  $\dot\alpha\cos\delta$ and is equivalent to
  $\mu^*_\alpha$.}, subscript $_{\rm DR3}$ represents
quantities at epoch $t_{\rm DR3}$,
superscript $^b$ represents TSB, and $\Delta$ means offset. Considering that the Gaia
measurements of systematic RVs are quite uncertain, we use Gaia DR3
 RVs to propagate astrometry instead of using them
to constrain reflex motion. 
\item {\bf Model astrometry of TSB at Hipparcos and Gaia DR3 reference
    epochs.} We model the TSB astrometry at the Hipparcos reference
  epoch $t_{\rm HIP}$ through linear propagation of state vectors in
  the Cartesian coordinate system as follows:
  \begin{eqnarray}
    \vec{r}_{\rm HIP}^b&=&\vec{r}^b_{\rm DR3}+\vec{v}^b_{\rm DR3}(t_{\rm HIP}-t_{\rm DR3})~,\\
    \vec{v}_{\rm HIP}^b&=&\vec{v}^b_{\rm DR3}~,
  \end{eqnarray}
  where $(\vec{r}_{\rm HIP}^b,\vec{v}_{\rm HIP}^b)$ and $(\vec{r}_{\rm DR3}^b,\vec{v}_{\rm DR3}^b)$ are respectively the
  state vectors, including location and velocity, of the TSB at the Hipparcos and Gaia DR3
  epochs. We first transform TSB astrometry from equatorial
  coordinate system, $\vec{\iota}^b_{\rm DR3}=(\alpha^b_{\rm
    DR3},\delta^b_{\rm DR3},\varpi^b_{\rm DR3},\mu^b_{\alpha {\rm
      DR3}},\mu^b_{\delta {\rm DR3}},{\rm RV}_{\rm DR3})$, to Cartesian coordinate system to get state
  vector at Gaia DR3 epoch, and then propagate the vector to the
  Hipparcos epoch, and then transform the new vector back to the
  astrometry at the Hipparcos epoch, $\vec{\iota}^b_{\rm HIP}$. The
  whole process is: equatorial state vector at $t_{\rm DR3}$
  $\rightarrow$ Cartesian state vector at $t_{\rm DR3}$ $\rightarrow$
  linear propagation of Cartesian state vector to $t_{\rm HIP}$
  $\rightarrow$ equatorial state vector at $t_{\rm HIP}$. By propagating state vectors in Cartesian coordinate system
  instead of spherical coordinate system, this approach completely solve the problem of perspective acceleration. The transformation between different coordinate systems is described in
  \cite{lindegren12} and \cite{feng19c}.
\item {\bf Simulate Gaia abscissae using GOST.} Instead of
  simulating the Gaia epoch data precisely by considering relativistic
  effects, perspective acceleration, and instrumental effects
  \citep{lindegren12}, we simulate the Gaia abscissae (or along-scan
  coordinates) by only considering linear motion of the TSB in the equatorial
  coordinate system as well as the target reflex motion \footnote{It is actually the motion of system photocenter rather than the mass center of the target star. Considering that the companion is far smaller than its host star in this work, the two centers are almost identical.}. This
  is justified because the various effects are independent
  of reflex motion, and can be estimated and subtracted from the data
  a priori. 

  We simulate the position of the target at GOST epoch $t_j$ relative
  to the Gaia DR3 reference position by adding the stellar reflex motion (denoted by superscript $^r$) onto the TSB motion,
    \begin{eqnarray}                                                                                         
    \Delta \alpha_{*i}&=&\Delta \alpha^b_{*\rm DR3}+\mu_{\alpha {\rm DR3}}^b(t_i-t_{\rm DR3})+ \Delta\alpha^r_{*i}~, \\                                       
    \Delta \delta_{i}&=&\Delta\delta^b_{\rm DR3} + \mu_{\delta {\rm DR3}}^b(t_i-t_{\rm DR3}) + \Delta\delta^r_i~,                                                
    \end{eqnarray}
where $\Delta \alpha^b_{*\rm DR3}=(\alpha^b_{\rm DR3}-\alpha_{\rm
  DR3})\cos\delta^b_{\rm DR3}$, and $\Delta \delta^b_{\rm DR3}=\delta^b_{\rm DR3}-\delta_{\rm DR3}$.
Because the reflex motion caused by cold Jupiters is insignificant
compared with barycentric motion, the parallax at epoch $t_i$ is approximately
$\varpi_i=\varpi^b_i+\Delta\varpi^r_i=\varpi^b_{\rm DR3}+\Delta\varpi^r_i\approx
\varpi^b_{\rm DR3}$, and the systematic radial velocity is
approximately ${\rm RV}_i={\rm RV}^b_i+\Delta{\rm RV}^r={\rm RV}^b_{\rm DR3}+\Delta{\rm RV}^r\approx {\rm RV}_{\rm DR3}$.

  Considering that Gaia pixels in along-scan (AL) direction are much
  smaller than that in the cross-scan direction, we only model the
  along-scan position of the target (or abscissa; $\eta_i$). Considering the parallax caused by Gaia's heliocentric
  motion, abscissa is modeled by projecting the motion of the target onto the AL direction using
  \begin{equation}
    \eta_i= \Delta
    \alpha_{*i}\sin\psi_i+\Delta\delta_i\cos\psi_i+\varpi^b_{\rm DR3}f^{\rm AL}_i~.
  \end{equation}

\item {\bf Fit a five-parameter model to synthetic Gaia abscissae.}
  Considering that binary solution is only applied to a small
  fraction of DR3 targets and most planet-induced reflex motion is
  not yet available in the Gaia non-single star catalog \citep{gaia22}, we model the simulated abscissae using a five-parameter model as follows:
  \begin{eqnarray}
    \hat\eta_i&=&\Delta\alpha_{*{\rm
                  DR3}}^l\sin\psi_i+\Delta\delta_{\rm
                  DR3}^l\cos\psi_i+\hat\varpi_{\rm DR3} f^{\rm AL}_i~,\label{eq:abs}\\
    \Delta \alpha_{*{\rm DR3}}^l&=&(\hat\alpha_{\rm DR3}-\alpha_{\rm
                             DR3})\cos\hat\delta_{\rm DR3}+\hat\mu_{\alpha
                     {\rm DR3}}(t_i-t_{\rm DR3})~,\\
    \Delta \delta_{\rm DR3}^l&=&(\hat\delta_{\rm DR3}-\delta_{\rm DR3})+\hat\mu_{\delta
                         {\rm DR3}}(t_i-t_{\rm DR3})~,
  \end{eqnarray}
where $\hat\vec\iota_{\rm DR3}=(\hat\alpha_{\rm DR3}, \hat\delta_{\rm
  DR3}, \hat\varpi_{\rm DR3}, \hat\mu_{\alpha {\rm
    DR3}},\hat\mu_{\delta {\rm DR3}})$ is the set of model parameters at $t_{\rm
  DR3}$, $f_i^{\rm AL}$ is the along-scan parallax factor, and $\psi_i$ is the scan angle at epoch $t_i$. This scan
angle is the complementary angle of $\psi$ in the new Hipparcos
IAD \citep{leeuwen07,brandt21,holl22}, and thus $\psi$ will be replaced by
$\pi/2-\psi$ when modeling Hipparcos IAD. The above modeling of Gaia
DR3 can be applied to Gaia DR2 by changing the subscript
$_{\rm DR3}$ into $_{\rm DR2}$. Given the limited information
available through GOST, we are unable to reconstruct the uncertainties
of individual observations and which epochs are actually used in
producing the catalogs. As long as the astrometric uncertainties and rejected
epochs are not significantly time-dependent, it is reasonable to assume that all GOST
epochs are used in the astrometric solution of Gaia and all abscissae
have the same uncertainty. Under this assumption, we fit the
five-parameter model shown in eq. \ref{eq:abs} to the simulated
abscissae ($\eta_i$) for Gaia DR2 and DR3 through linear regression.
\item {\bf Calculate the likelihood for Gaia DR2 and DR3.} To avoid
  numerical errors, the catalog astrometry at $t_i$ relative to the
  Gaia DR3 epoch is defined as $\Delta\vec{\iota}_i\equiv
  (\Delta\alpha_{*i},\Delta\delta_i,\Delta\varpi_i,\Delta\mu_{\alpha
    i},\Delta\mu_{\delta i})=((\alpha_i-\alpha_{\rm
    DR3})\cos\delta_i,\delta_i-\delta_{\rm DR3},\varpi_i-\varpi_{\rm
    DR3},\mu_{\alpha i}-\mu_{\alpha {\rm DR3}},\mu_{\delta
    i}-\mu_{\delta {\rm DR3}})$. The fitted astrometry for epoch $t_i$ is
  $\Delta\hat\vec{\iota}_i$. The likelihood for Gaia DR2 and DR3 is
  \begin{equation}
    \mathcal{L}_{\rm gaia}=\prod_{i=1}^{N_{\rm DR}}(2\pi)^{-5/2}({\rm det}\Sigma'_i)^{-\frac{1}{2}}e^{-\frac{1}{2}(\Delta\hat\vec{\iota}_i
-\Delta\vec{\iota}_i)^T[\Sigma_i(1+J_i)]^{-1}(\Delta\hat\vec{\iota}_i-\Delta\vec{\iota}_i)}~,   
\end{equation}
where $N_{\rm DR}$ is the number of data releases used in the
analyses, and $\Sigma_i(1+J_i)$ is the jitter-corrected
covariance for the five-parameter solutions of Gaia DRs, $i=1$ and 2
respresent Gaia DR2 and DR3, respectively. In this work, we only
use DR2 and DR3 and thus $N_{\rm DR}=2$.
\item {\bf Calculate the likelihood for Hipparcos intermediate data.}
  We model the abscissae of Hipparcos ($\xi$) by adding the reflex
  motion onto the linear model shown in eq. \ref{eq:abs}, and calculate the likelihood as follows:
  \begin{equation}
    \mathcal{L}_{\rm hip}=\prod\limits_{i=1}^{N_{\rm epoch}}\frac{1}{\sqrt{2\pi(\sigma_i^2+J_{\rm hip}^2)}}e^{-\frac{(\hat\xi_i
-\xi_i)^2}{2(\sigma_i^2+J_{\rm hip}^2)}},
\end{equation}
where $N_{\rm epoch}$ is the total number of epochs of Hipparcos IAD. 
\end{itemize}
To obtain the total likelihood ($\mathcal{L}=\mathcal{L}_{\rm RV}\cdot
\mathcal{L}_{\rm hip}\cdot\mathcal{L}_{\rm gaia}$), we derive the likelihoods for the Gaia and Hipparcos data through the above steps,
and calculate the likelihood for the RV data ($\mathcal{L}_{\rm RV}$) following \cite{feng17b}. We adopt log uniform priors for
time-scale parameters such as period and correlation time scale in the
moving average (MA) model \citep{feng17b} and uniform priors for other
parameters. Finally, we infer the orbital parameters by sampling the
posterior through adaptive and parallel Markov Chain Monte Carlo (MCMC)
developed by \cite{haario01} and \cite{feng19a}.

While packages like \texttt{orbitize} \citep{blunt20},
  \texttt{FORECAST} \citep{bonavita22}, \texttt{BINARYS}
  \citep{leclerc23}, and \texttt{kepmodel} \citep{delisle22} have been developed to analyze imaging, RV, and astrometric data, they are not primarily designed for analyzing Gaia catalog data as performed by \texttt{orvara} \citep{brandt21b} and \texttt{htof} \citep{brandt21}. Compared with \texttt{orvara} and \texttt{htof}, our method has the following features:
  \begin{itemize}
  \item we simultaneously optimize all model parameters through MCMC
    posterior sampling;
  \item we utilize multiple Gaia data releases by fitting multiple
    five-parameter models to Gaia epoch data simulated by
    GOST\footnote{Although \texttt{htof} also simulates Gaia epoch data
      using GOST, it does not utilize multiple Gaia DRs.};
  \item instead of conducting calibration a priori, we employ jitters and offsets to model systematics a posteriori;
  \item we model time-correlated noise in the RV data. 
  \end{itemize}

While the use of multiple Gaia DRs does not significantly improve the constraint on long-period orbits by increasing the temporal baseline between Gaia and Hipparcos, it does provide additional information about the raw abscissae. This, in turn, leads to stronger orbital constraints at Gaia epochs. In other words, Gaia DR2 provides a 5D astrometric data point, and when combined with the 5D data point from Gaia DR3, the orbital constraint becomes stronger. The additional information obtained from multiple DRs enhances the accuracy and reliability of our orbital solutions.

Through sensitivity tests, we find that our orbital solutions
are not strongly sensitive to whether or not we calibrate frame
rotation and zero-point parallax a priori by adopting values from
previous studies \citep{brandt18, kervella19, lindegren20}. It is
important to note that there is uncertainty in the estimation of these
calibration parameters, as indicated by studies such as
\cite{brandt18} and \cite{lindegren20}. Additionally, uncertainties
can be amplified during the transformation from the Gaia frame to the
Hipparcos frame. Considering these factors, it is more appropriate to
consider astrometric systematics on a case-by-case basis, taking into
account individual target characteristics. These issues have been
discussed in our previous work \citep{feng21}.

\begin{figure}
  \centering
  \includegraphics[scale=0.4]{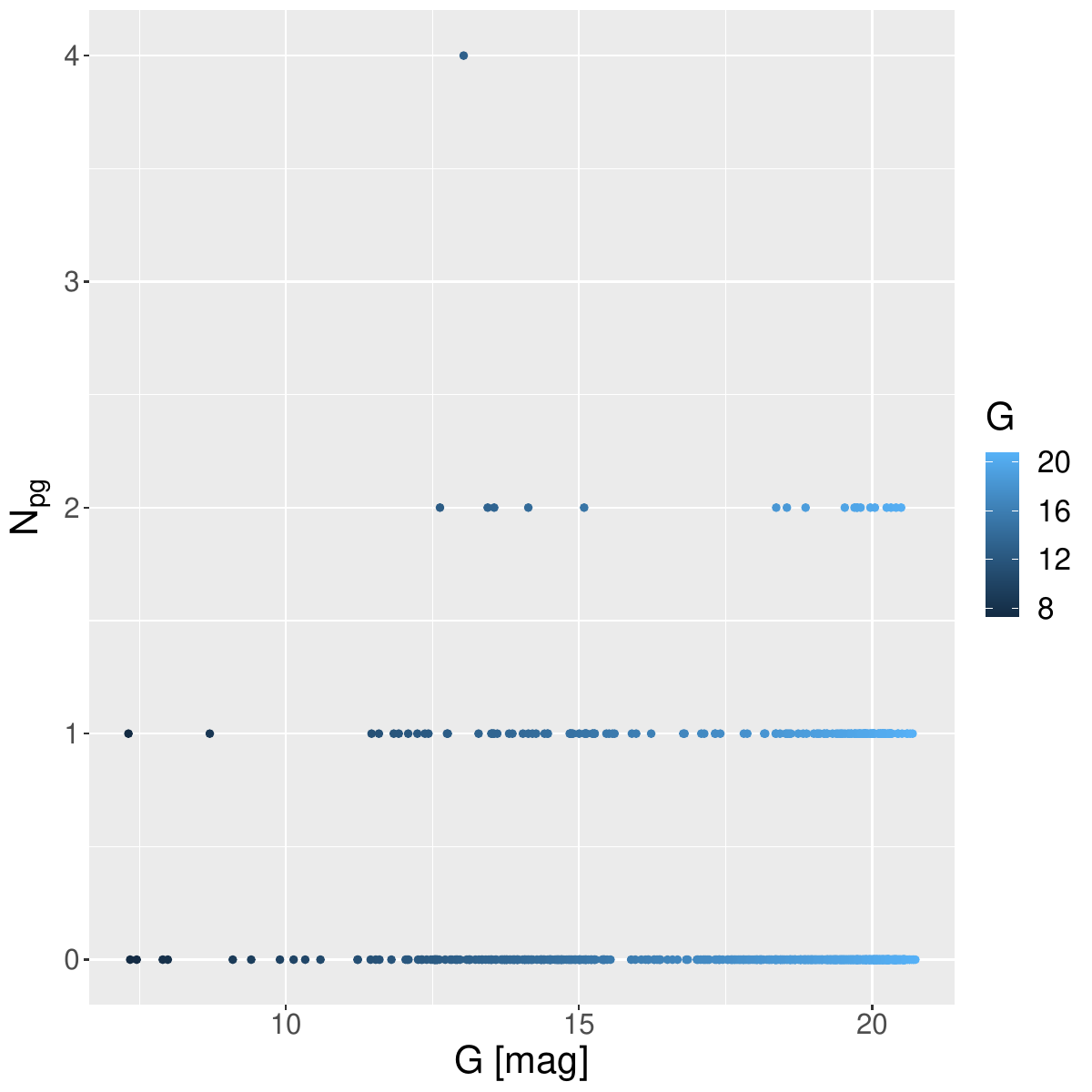}
    \includegraphics[scale=0.4]{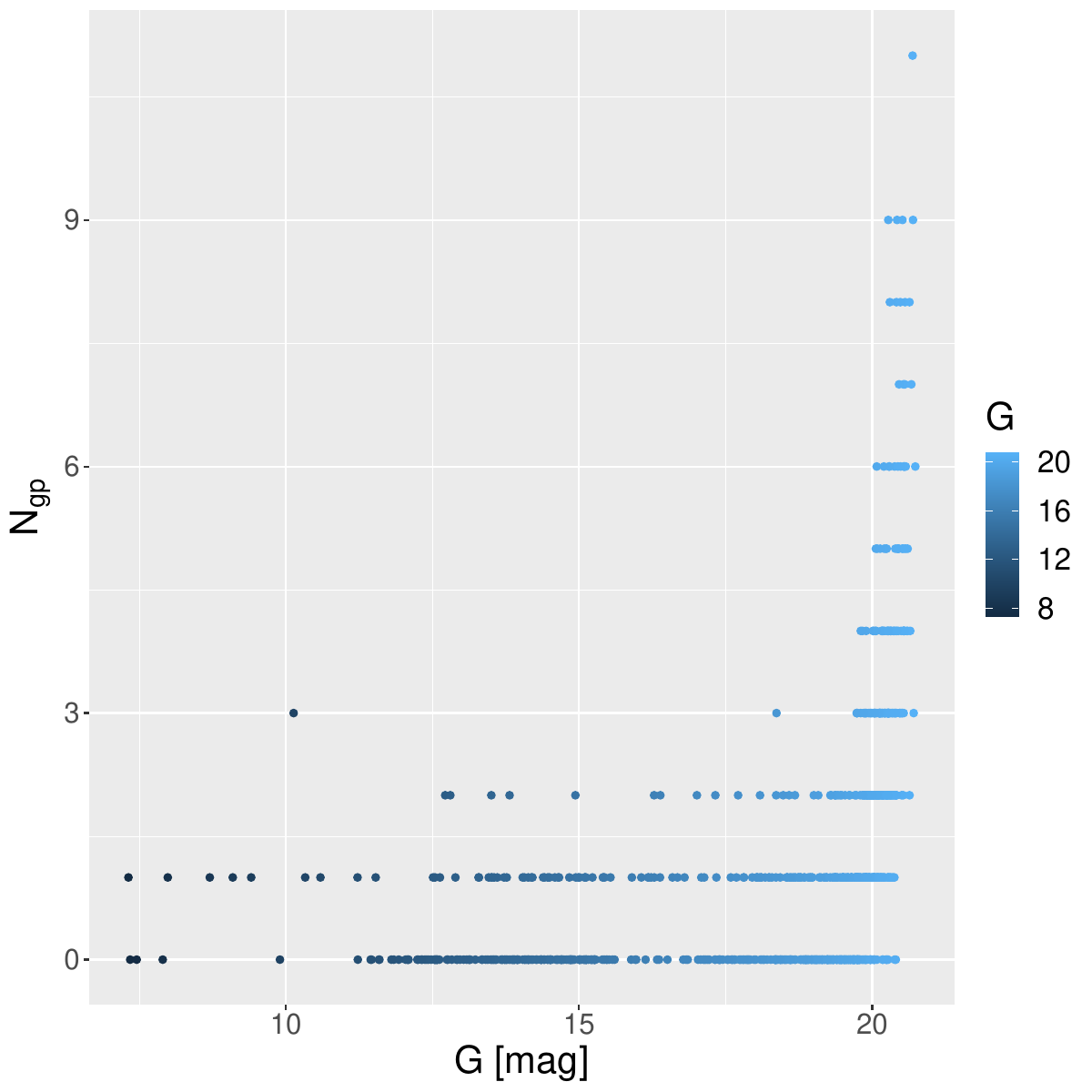}
  \caption{Distribution of 1000 stars over the G-band magnitude
    and the number of mismatched epochs between GOST and the GDR3
    epoch-photometry catalog. The left panel is for $N_{\rm pg}$ while
    the right panel is for $N_{\rm gp}$.}
  \label{fig:epoch_comp}
\end{figure}

To validate the accuracy of GOST emulations, we perform a comparison
between the astrometric epochs generated by GOST and the G band
transit times provided in Gaia Data Release 3 (GDR3) for a randomly
selected sample of 1000 stars that have epoch photometry. We assess
the number of mismatched epochs between GOST and the GDR3
epoch-photometry catalog, as well as the distribution of these
mismatches. We define $N_{\rm gp}$ as the number of epochs predicted by GOST but not present in the GDR3 epoch-photometry catalog, and $N_{\rm pg}$ as the number of epochs present in the photometry catalog but not predicted by GOST. The distribution of the sample over G magnitude and the number of mismatched epochs is depicted in Figure \ref{fig:epoch_comp}. Notably, it is apparent that for bright stars with G magnitudes less than 10, the number of photometric epochs mismatched between GOST and GDR3 is at most 1.

To quantify the fraction of missing epochs relative to the total
photometric epochs, we calculate $\eta_{\rm gp} = N_{\rm gp}/N_p$ and
$\eta_{\rm pg} = N_{\rm pg}/N_p$, where $N_p$ represents the total
number of G band photometric transits. The median of $N_p$ is 22. The
mean values of $\eta_{\rm gp}$ and $\eta_{\rm pg}$ for bright stars
are found to be 2.1\% and 0.7\%, respectively. For all stars, the mean values of $\eta_{\rm gp}$ and $\eta_{\rm pg}$ are 7.0\% and 0.9\%, respectively. This analysis serves as a validation of the accuracy of GOST emulations, further supporting the reliability of our approach.

\section{Results}\label{sec:results}
\begin{figure*}
  \centering
  \includegraphics[scale=0.45]{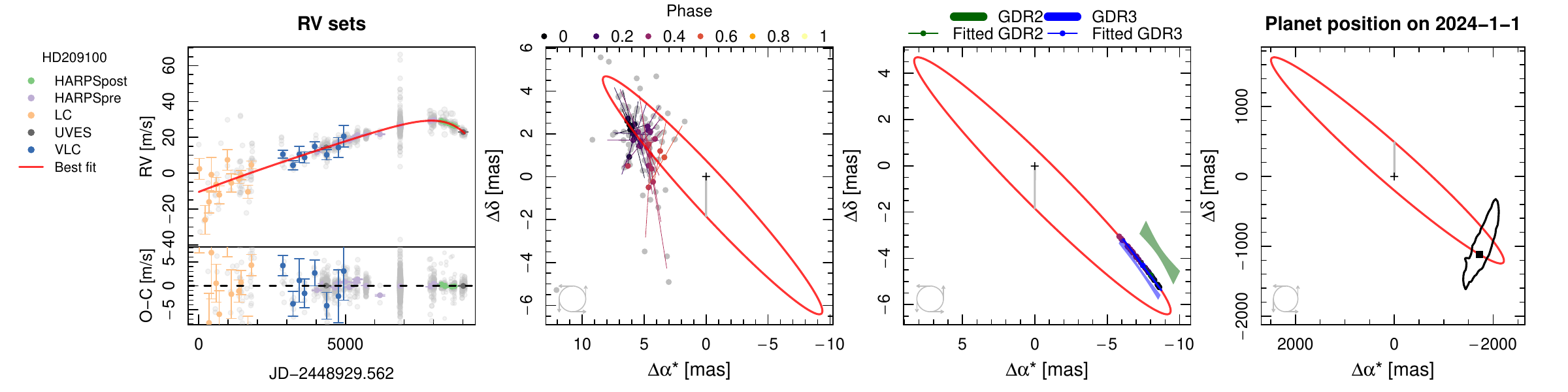}
  \includegraphics[scale=0.45]{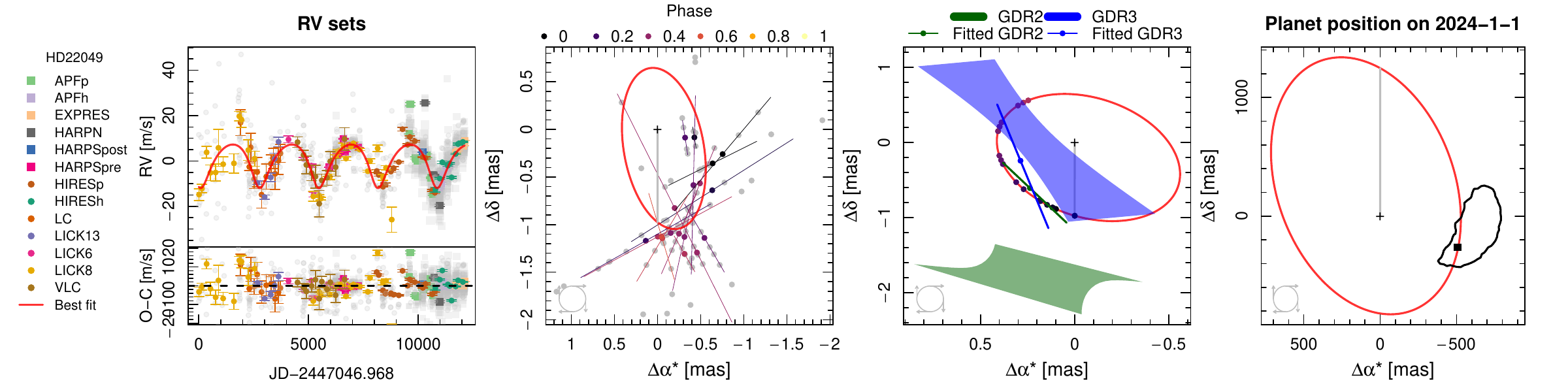}
  \caption{Optimal orbital solutions for \ind b (top) and \eri b (bottom). The
    panels from the left to the right respectively show the best fit to RV, Hipparcos IAD, and
    Gaia GOST data, and the predicted planetary position on January
    1st, 2024. The first column
    shows the binned RV data sets encoded by colors and shapes. For
    optimal visualization, each RV set is binned with a 100\,d time window while the un-binned RVs and residuals are shown in grey. The second column shows the
    post-fit Hipparcos abscissa residual projected along the R.A. and
    decl. directions. The multiple measurements for each epoch are binned to
    present the binned data encoded by colors. The darker colors encode
    earlier phases while the brighter colors represent later
    phases. The directions of the error bars indicate the
    along-scan direction. The third column shows the optimal fit to the Gaia GOST data
    and the comparison between best-fit and catalog proper motions and
    positions at Gaia DR2 (GDR2) and GDR3 reference epochs. The
    shaded regions represent the uncertainty of position and proper
    motion. The dot and slope of each line respectively represent the best-fit
    position and proper motion offsets induced by the reflex motion at certain
    reference epoch. The fourth column shows the predicted planet position on January 1st, 2024. The 1-$\sigma$ contour line is
  shown to indicate the prediction uncertainty. }
\label{fig:fit}
\end{figure*}

\begin{figure*}
  \centering
  \includegraphics[scale=0.4]{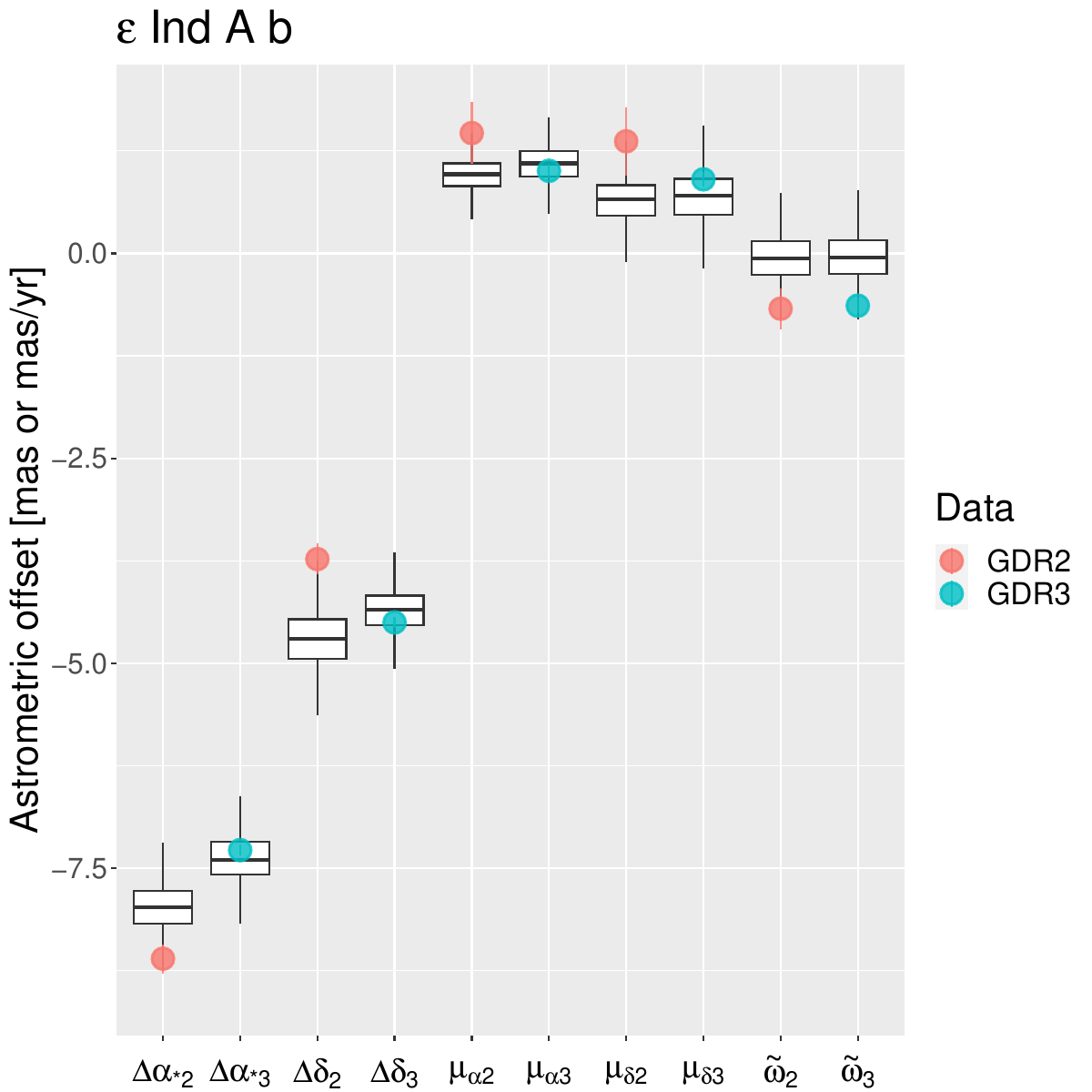}
    \includegraphics[scale=0.4]{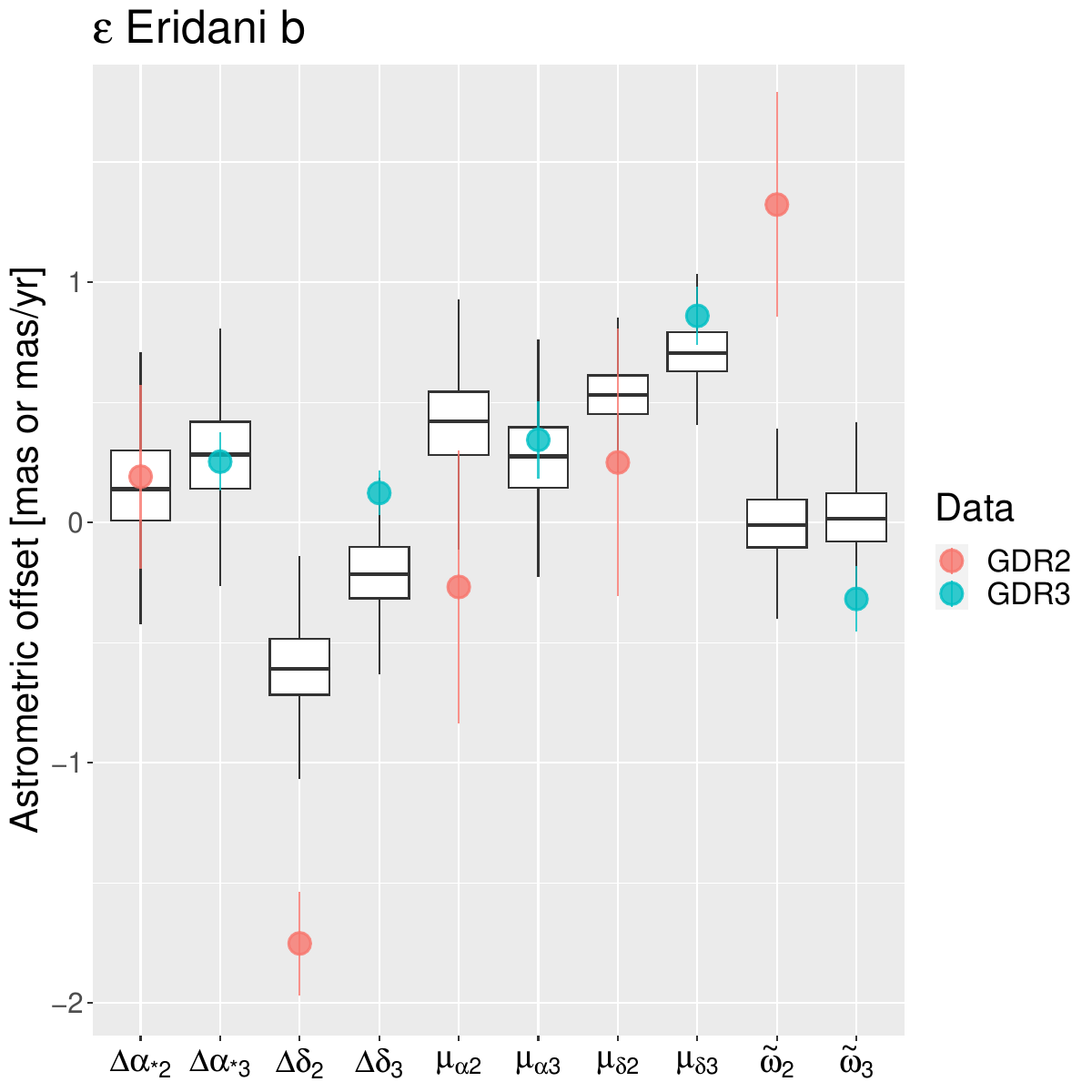}
  \caption{Model fit to the five-parameter astrometry of GDR2 and
    GDR3 for \ind b (left) and \eri b (right). In both cases, the barycentric
    astrometry is subtracted from the five-parameter solutions for
    both the data (represented by a dot with error bar) and the model
    prediction (represented by a boxplot). The subscripts of the labels on the x-axes indicate the Gaia data release number.}
\label{fig:5fit}
\end{figure*}

We find the optimal orbital solutions for \ind b and \eri b based on
the MCMC samplings of posteriors and show the solutions in
Fig. \ref{fig:fit} and the posterior distributions of orbital
parameters in Fig. \ref{fig:HD209100_corner} and
Fig. \ref{fig:HD22049_corner}. To optimize the visualization of
Fig. \ref{fig:fit}, we project the
synthetic abscissae along the R.A. and decl. direction, encode the orbital phase
with colors, and represent the orbital direction using circles with
arrows. In the panels for Gaia synthetic abscissae (third column), we use segments
and shaded regions to visualize the catalog astrometry
($\Delta\vec\iota-\Delta\hat\vec\iota^b$) and fitted astrometry ($\Delta\hat\vec\iota-\Delta\hat\vec\iota^b$) after subtracting
the astrometry of TSB, respectively. The center of the segment is
determined by the R.A. and decl. relative to the TSB, the slope is the
ratio of proper motion offsets in the decl. and R.A directions, and
the length is equal to the proper motion offset multiplied by the time
span of Gaia DR2 or DR3. The fitted astrometry is determined through a five-parameter linear fit to the synthetic data, which is represented by colorful dots in the panels of the third column of Fig. \ref{fig:fit}.  We also predict the position of \ind b and
\eri b on January 1st, 2024. Their angular separations are 2.1$\pm$0.1\as and
0.64$\pm$0.10\as, and position angles are $243\pm 14$\deg and $258\pm
30$\deg, respectively. In addition, we present the fit to the
five-parameter astrometry of Gaia DR2 and DR3 in
Fig. \ref{fig:5fit}. It is apparent that the deviation caused by \ind
b from the barycentric astrometry is more pronounced compared to that
induced by \eri b. The model fit consistently remains within 1-$\sigma$ of the five-parameter catalog astrometry. 

  
The inferred parameters are presented in Table \ref{tab:solution},
\ref{tab:other_HD209100}, and \ref{tab:other_HD22049}. In
the table, the following orbital parameters are directly inferred
through posterior sampling: the orbital period ($P$), RV semi-amplitude ($K$), eccentricity
($e$), argument of periastron ($\omega$) of the stellar reflex
motion\footnote{Note that the argument of periastron for the planetary
  orbit is $\omega_p=\omega+\pi$.}, inclination ($I$), longitude of
ascending node ($\Omega$), and mean anomaly at the minimum epoch of RV
data ($M_0$). The table also presents the astrometric offsets, which
need to be subtracted from the catalog astrometry of Gaia DR3 to
derive the TSB astrometry. The bottom three rows present the derived
parameters, including the semi-major axis of the planet-star binary
orbit ($a$), planet mass ($m_p$), and the epoch when a planet crosses
through its periastron ($T_p$).

\begin{table}
\caption{Parameters for \ind b and \eri b.}
\label{tab:solution}
\begin{center}
\begin{tabular}{lp{.10\textwidth}lllllr}
\hline\hline
Parameter$^a$ &Unit & Meaning& \ind b & \eri b & Prior$^c$ & Minimum & Maximum\\
  \hline
  $P$&day&Orbital period&$15676.48_{-1492.59}^{+2331.54}$&$2688.60_{-16.51}^{+16.17}$ & Log-Uniform&-1&16\\
$K$&\ms&RV semi-amplitude&$31.69_{-4.25}^{+4.55}$&$9.98_{-0.38}^{+0.43}$ & Uniform&$10^{-6}$&$10^{6}$\\
$e$&---&Eccentricity&$0.42_{-0.04}^{+0.04}$&$0.26_{-0.04}^{+0.04}$ & Uniform&0&1\\
$\omega$$^{b}$&deg&Argument of periapsis&$98.73_{-11.46}^{+9.46}$&$166.48_{-6.66}^{+6.63}$ & Uniform&0&2$\pi$\\
$I$&deg&Inclination&$84.41_{-9.94}^{+9.36}$&$130.60_{-12.62}^{+9.53}$ & CosI-Uniform&-1&1\\
$\Omega$&deg&Longitude of ascending node&$243.38_{-13.41}^{+14.36}$&$206.07_{-17.48}^{+15.14}$ & Uniform&0&2$\pi$\\
$M_0$&deg&Mean anomaly at the reference epoch&$127.64_{-24.09}^{+31.05}$&$352.80_{-10.00}^{+10.80}$ & Uniform&0&2$\pi$\\\hline
$\Delta \alpha$&mas&$\alpha$ offset&$-7.51_{-1.72}^{+1.59}$&$0.26_{-0.20}^{+0.19}$ & Uniform&$-10^6$&$10^6$\\
$\Delta \delta$&mas&$\delta$ offset&$-3.83_{-2.16}^{+2.40}$&$-0.04_{-0.20}^{+0.18}$ & Uniform&$-10^6$&$10^6$\\
$\Delta \varpi$&mas&$\varpi$ offset&$-0.59_{-0.33}^{+0.28}$&$-0.31_{-0.15}^{+0.14}$ & Uniform&$-10^6$&$10^6$\\
$\Delta \mu^*_\alpha$&mas\,yr$^{-1}$&$\mu^*_\alpha$ offset &$1.00_{-0.08}^{+0.11}$&$0.36_{-0.02}^{+0.02}$ & Uniform&$-10^6$&$10^6$\\
$\Delta \mu_\delta$&mas\,yr$^{-1}$&$\mu_\delta$ offset&$0.95_{-0.19}^{+0.21}$&$0.85_{-0.01}^{+0.01}$ & Uniform&$-10^6$&$10^6$\\\hline
$P$&yr&Orbital period &$42.92_{-4.09}^{+6.38}$&$7.36_{-0.05}^{+0.04}$ & ---&---&---\\
$a$&au&Semi-major axis &$11.08_{-0.74}^{+1.07}$&$3.53_{-0.06}^{+0.06}$ & ---&---&---\\
$m_p$&$M_{\rm Jup}$&Planet mass &$2.96_{-0.38}^{+0.41}$&$0.76_{-0.11}^{+0.14}$ & ---&---&---\\
$T_p-2400000$&JD&Periapsis epoch&$43293.68_{-1455.27}^{+1202.96}$&$44411.54_{-81.95}^{+76.60}$ & ---&---&---\\
\hline
\multicolumn{8}{p{1\textwidth}}{\footnotesize$^a$ The first 12 rows
  show parameters that are inferred directly through MCMC posterior
  sampling, while the last five rows show the parameters derived from
  the directly sampled parameters. The semi-major axis $a$ and planet
  mass $m_p$ are derived from the orbital parameters by adopting a stellar mass of 0.81$\pm$0.04\msun for \eri and 0.74$\pm$0.04\msun for \ind from the Gaia Final Luminosity Age Mass Estimator \citep{creevey22}. }\\
    \multicolumn{8}{p{1\textwidth}}{\footnotesize$^b$ This is the argument of periastron of stellar reflex motion and $\omega+\pi$ is the argument of periastron of planetary orbit.}\\
  \multicolumn{8}{p{1\textwidth}}{\footnotesize$^c$ The rightest three
  columns show the prior distribution and the corresponding minimum and maximum
  values for a parameter. ``Log-Uniform'' is the logarithmic uniform
  distribution, and ``CosI-Uniform'' is the uniform distribution over
  $\cos I$.}\\
 \end{tabular}
\end{center}
\end{table}

As the nearest cold Jupiters, \ind b and \eri b have been intensively
studied in the past. Here we compare our solutions with previous
ones. Based on combined analyses of RV, Hipparcos and
Gaia DR2, \cite{feng19b} determine a mass of
3.25$_{-0.65}^{+0.39}$\mj, a period of $45.20^{+5.74}_{-4.77}$\,yr,
and an eccentricity of $0.26_{-0.03}^{+0.07}$ for \ind b. Recently,
\cite{philipot23} analyze the RV, Hipparcos and Gaia early data
release 3 (EDR3; \citealt{gaia22}) data using the HTOF package
\citep{brandt21}, and estimate a mass of $3.0\pm 0.1$\mj, a period of
$29.93_{-0.62}^{+0.73}$\,yr, and an eccentricity of $0.48\pm
0.01$. In addition to the Gaia EDR3 (equivalent to DR3 for
five-parameter solutions) data used by \cite{philipot23},
we use Gaia DR2 to constrain the orbit. We find a mass of $2.96_{-0.38}^{+0.41}$\mj, a period of $42.92^{+6.38}_{-4.09}$\,yr,
and an eccentricity of $0.42_{-0.04}^{+0.04}$. The orbital period
estimated in this work and in \cite{feng19b} is significantly longer
than the one given by \cite{philipot23} probably due to the following
reasons: (1) we optimize all parameters simultaneously instead of
  marginalizing some of them\footnote{As an example, \texttt{orvara} performs marginalization of the RV offsets before fitting the orbital parameters. It assumes that there is no correlation between the offset and the orbital period.}; (2) as suggested by \cite{zechmeister13}, we subtract the
  perspective acceleration (about 1.8\msyr) from both the CES LC and VLC sets; (3) we
  model the time-correlated RV noise as well as instrument-dependent
  jitters \citep{feng19b}; (4) we use astrometric data from both Gaia
  DR2 and DR3. A detailed comparison of our methodology with others is given by \cite{feng21}.

Although the existence of a cold Jupiter around \eri was disputed
due to consideration of stellar activity cycles in RV analyses
\citep{hatzes00,anglada12}, \eri b has been gradually confirmed by combined
analyses of RV, astrometry, and imaging data
\citep{benedict17,mawet19,llop21,benedict22,roettenbacher22}. Based on RV constraint and
direct imaging upper limit of \eri b, \cite{mawet19} determine a mass
of $0.78_{-0.12}^{+0.38}$\mj, a period of $7.37\pm 0.07$\,yr, and an
eccentricity of $0.07_{-0.05}^{+0.06}$. \cite{llop21} analyze RV,
absolute astrometry from Hipparcos and Gaia DR2, and imaging data, and
estimate a mass of $0.66_{-0.09}^{+0.12}$\mj, a period of
$2671_{-23}^{+17}$\,d, and an eccentricity of
$0.055_{-0.039}^{+0.067}$. With additional RV data from EXPRES,
\cite{roettenbacher22} get a solution consistent with that given by
\cite{mawet19}. Using only the astrometric data obtained by the Fine
Guidance Sensor 1r of HST, \cite{benedict22} determines a mass of
$0.63_{-0.04}^{+0.12}$\mj, a period of 2775$\pm$5\,d, and an
eccentricity of 0.16$\pm$0.01. In this work, the combined analyses of
RV, Hipparcos, Gaia DR2 and DR3 astrometry determine a mass of
$0.76_{-0.11}^{+0.14}$\mj, a period of $2688.60_{-16.51}^{+16.17}$\,d,
and an eccentricity of 0.26$\pm$0.04. Modeling the time-correlated RV
noise using the MA model, our solution gives an eccentricity significantly higher than the value given by
\cite{mawet19}, \cite{llop21}, and \cite{roettenbacher22}, who use Gaussian processes to model
time-correlated RV noise. While Gaussian process interprets the
extra RV variations at the peaks and troughs of the Keplerian signal as
time-correlated noise (see the bottom left panel of Fig. \ref{fig:fit}
in this paper and fig. 2 in \citealt{roettenbacher22}), we interpret them as part of the signal because
these excessive variations always amplify the peaks and troughs at the
corresponding epochs with a similar periodicity. In previous studies where Gaussian process is not employed in RV modeling, high eccentricities have typically been estimated, such as $0.608\pm 0.041$\citep{hatzes00}, $0.25\pm0.23$
  \citep{butler06}, and $0.40\pm0.11$\citep{anglada12}. Furthermore, astrometry-only analyses have also resulted in high-eccentricity solutions, as demonstrated by previous studies such as $e=0.702\pm0.039$ \citep{benedict06} and $e=0.16\pm 0.01$ \citep{benedict22}. Therefore, it is plausible to conclude that the orbital eccentricity of \eri b is significantly higher than zero, as indicated by the findings of this study. In our study, the inclination is
  $130.60_{-12.62}^{+9.53}$\deg. This is equivalent to an inclination of $49.40_{-9.53}^{+12.62}$\deg, which is calculated in the absence of knowledge regarding the longitude of the ascending node. This value
is largely consistent with the debris disk inclination ranging from
17\deg to 34\deg \citep{macGregor15,booth17} while \cite{llop21} determines a significantly high inclination of $78.81_{-22.41}^{+29.34}$\deg.

\section{Conclusion}\label{sec:conclusion}
Using GOST to emulate Gaia epoch data, we analyze both Gaia DR2 and
DR3 data in combination with Hipparcos and RV data to constrain the
orbits of the nearest Jupiters. For \ind b and
\eri b, the orbital periods are $42.92_{-4.09}^{+6.38}$\,yr and
$7.36_{-0.05}^{+0.04}$\,yr, the eccentricities are $0.42_{-0.04}^{+0.04}$ and $0.26_{-0.04}^{+0.04}$, and the masses are $2.96_{-0.38}^{+0.41}$\mj and $0.76_{-0.11}^{+0.14}$\mj,
respectively. It is the first time that both Gaia DR2 and DR3 catalog data are modeled and analyzed by emulating the Gaia
epoch data with GOST. Compared with previous studies, our approach avoids approximating
instantaneous astrometry by catalog astrometry at the reference epoch,
and thus enable robust constraint of orbits with period comparable or
shorter than the DR2 and DR3 observation time span. 

The orbital period of \ind b in this work is much longer than the
value given by \cite{philipot23}, but is consistent with the solution
based on combined analyses of RV and Gaia DR2 \citep{feng19b}. While we estimate an orbital
eccentricity of \eri b much higher than value given by \cite{mawet19},
\cite{llop21} and \cite{roettenbacher22} who use Gaussian
process to model time-correlated noise, our solution is largely
consistent with previous studies
\citep{hatzes00,anglada12,benedict06,benedict22}, which don't use
Gaussian process to model stellar activity. It is possible that Gaussian
process may interpret part of the Keplerian signal as time-correlated
noise \citep{feng16,ribas18}.

The combined RV and astrometry model presented in this work shares
similarities with the successful model used to detect and confirm
numerous sub-stellar and planetary companions in previous studies
\citep{feng19b,feng21,feng22}. The orbital solution obtained for \ind
using our new method aligns closely with the results determined in
\cite{feng19b}, which highlights the reliability of our approach in
detecting cold and massive companions. The upcoming imaging observations of \eri and \ind with JWST will provide further validation of our method, particularly in utilizing multiple Gaia DRs to identify cold Jupiters. While future Gaia DR4 will release epoch data, the modeling framework developed in our study offers a general methodology for incorporating both catalog data and epoch data to constrain orbital parameters. This approach can be extended to combined analyses of astrometric data from various sources, including Gaia, photographic plates \citep{cerny21}, Tycho-2 \citep{hog00}, and future space missions such as the China Space Station Telescope \citep{fu23} and the Nancy Roman Space Telescope \citep{yahalomi21}.

\section*{Acknowledgements}
This work is supported by Shanghai Jiao Tong University 2030
  Initiative. We would like to extend our sincere appreciation to the Scientific Editor and the anonymous referee for their valuable comments and feedback on our manuscript. We also thank Xianyu Tan for helpful discussion about
atmosphere models of our targets. This work has made use of data from
the European Space Agency (ESA) mission Gaia
(https://www.cosmos.esa.int/gaia), processed by the Gaia Data
Processing and Analysis Consortium (DPAC, https://www.cosmos.esa.int/web/gaia/dpac/consortium). Funding for the DPAC has been provided by national institutions, in particular the
institutions participating in the Gaia Multilateral Agreement. This
research has also made use of the services of the portal exoplanet.eu
of The Extrasolar Planets Encyclopaedia, the ESO Science Archive
Facility, NASA’s Astrophysics Data System Bibliographic Service, and
the SIMBAD database, operated at CDS, Strasbourg, France. We also
acknowledge the many years of technical support from the UCO/Lick
staff for the commissioning and operation of the APF facility atop
Mt. Hamilton. All analyses were performed using R Statistical Software (v4.0.0; R Core Team 2020).
This paper is partly based on observations collected at
the European Organisation for Astronomical Research in the Southern
Hemisphere under ESO programmes:
072.C-0488,072.C-0513,073.C-0784,074.C-0012,076.C-0878,077.C-0530,078.C-0833,079.C-0681,192.C-0852,60.A-9036.

\section*{Data Availability}
The new RV data are available in the appendix while the Gaia and Hipaprcos data are publicly available.



\bibliographystyle{mnras}
\bibliography{nm} 

\appendix
\section{Posterior distribution of orbital parameters}
\begin{figure*}
  \centering
  \includegraphics[scale=0.45]{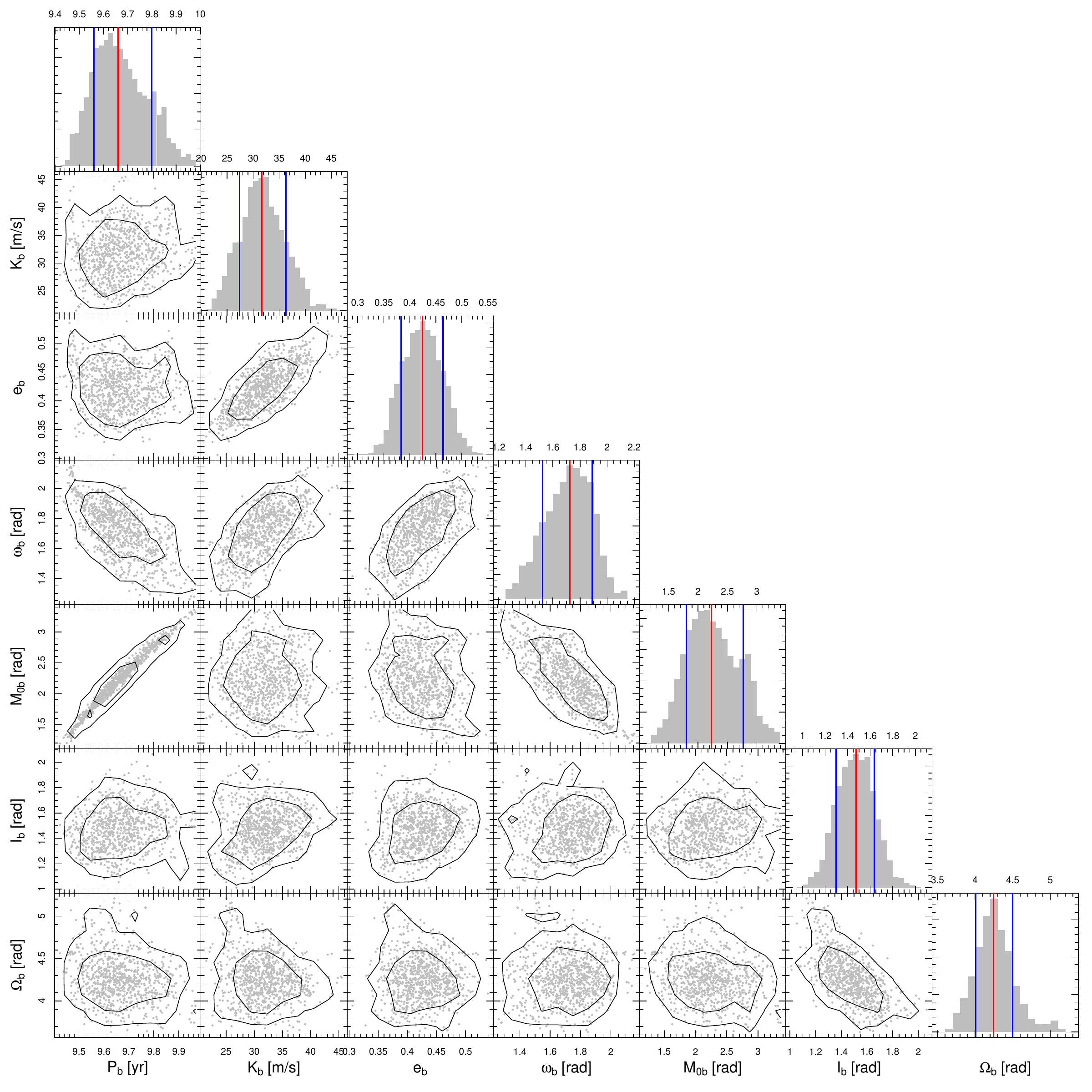}
  \caption{Corner plot showing the 1D and 2D posterior distribution of
    the orbital parameters for \ind b. In each histogram, the blue
    lines indicate the 1$\sigma$ confidence intervals, while the red
    line represents the median of the posterior distribution. The
    contour lines depict the 1$\sigma$ and 2$\sigma$ confidence
    intervals. Grey dots represent the posterior samples for each pair
    of parameters. It is important to note that $\omega_b$ denotes the argument of periastron for the stellar reflex motion, while the argument of periastron for the planetary orbit, $\omega_p$, is equal to $\omega_b+\pi$. Furthermore, $M_{0b}$ corresponds to the mean anomaly at JD 2447047.96844.}
  \label{fig:HD209100_corner}
\end{figure*}

\begin{figure*}
  \centering
  \includegraphics[scale=0.45]{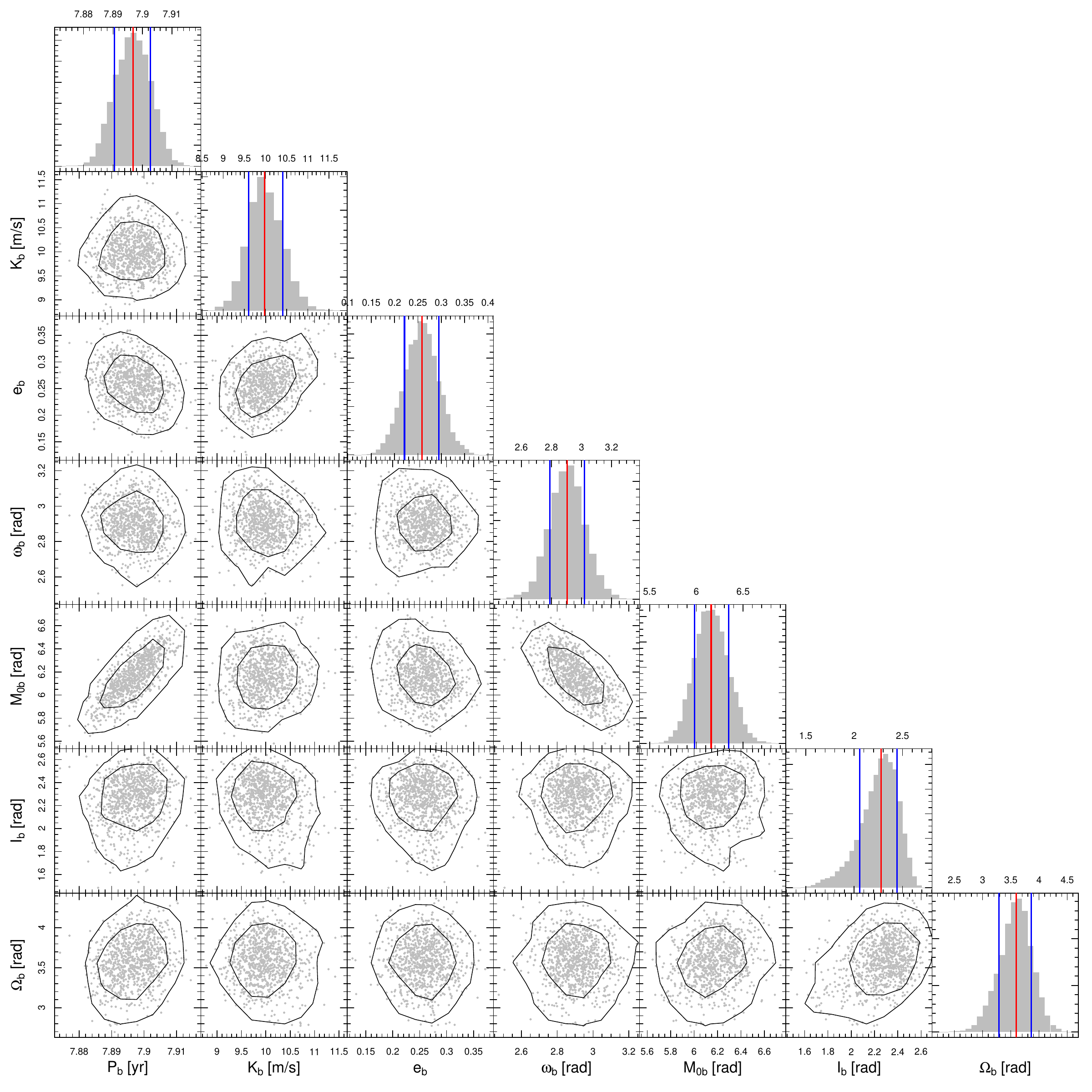}
  \caption{Similar to Fig. \ref{fig:HD209100_corner}, but for \eri
    b. $M_{0b}$ is the mean anomaly at JD 2448930.56223.}
  \label{fig:HD22049_corner}
\end{figure*}

\section{Other model parameters}
\begin{table}
\caption{Other parameters for \ind b.}
\label{tab:other_HD209100}
\begin{center}
\begin{tabular}{lp{.10\textwidth}llllr}
\hline\hline
Parameter$^a$ &Unit & Meaning& \ind b & Prior & Minimum & Maximum\\
\hline
$\gamma^{\rm HARPSpost}$&m\,s$^{-1}$&RV offset for HARPSpost&$-28.17_{-3.31}^{+2.97}$&Uniform&$-10^6$&$10^6$\\
$J^{\rm HARPSpost}$&m\,s$^{-1}$&RV jitter for HARPSpost&$1.11_{-0.08}^{+0.09}$&Uniform&0&$10^6$\\
$w_{1}^{\rm HARPSpost}$&---&Amplitude of component 1 of MA(1) for HARPSpost&$0.97_{-0.02}^{+0.02}$&Uniform&-1&1\\
ln$\tau^{\rm HARPSpost}$&---&Logarithmic time scale of MA(1) for HARPSpost&$1.01_{-0.21}^{+0.21}$&Uniform&-12&12\\
$\gamma^{\rm HARPSpre}$&m\,s$^{-1}$&RV offset for HARPSpre&$-25.62_{-3.33}^{+3.01}$&Uniform&$-10^6$&$10^6$\\
$J^{\rm HARPSpre}$&m\,s$^{-1}$&RV jitter for HARPSpre&$1.00_{-0.01}^{+0.01}$&Uniform&0&$10^6$\\
$w_{1}^{\rm HARPSpre}$&---&Amplitude of component 1 of MA(5) for HARPSpre&$0.40_{-0.01}^{+0.01}$&Uniform&-1&1\\
$w_{2}^{\rm HARPSpre}$&---&Amplitude of component 2 of MA(5) for HARPSpre&$0.25_{-0.02}^{+0.01}$&Uniform&-1&1\\
$w_{3}^{\rm HARPSpre}$&---&Amplitude of component 3 of MA(5) for HARPSpre&$0.15_{-0.02}^{+0.01}$&Uniform&-1&1\\
$w_{4}^{\rm HARPSpre}$&---&Amplitude of component 4 of MA(5) for HARPSpre&$0.06_{-0.01}^{+0.01}$&Uniform&-1&1\\
$w_{5}^{\rm HARPSpre}$&---&Amplitude of component 5 of MA(5) for HARPSpre&$0.12_{-0.01}^{+0.01}$&Uniform&-1&1\\
ln$\tau^{\rm HARPSpre}$&---&Logarithmic time scale of MA(5) for HARPSpre&$0.16_{-0.12}^{+0.11}$&Uniform&-12&12\\
$\gamma^{\rm LC}$&m\,s$^{-1}$&RV offset for LC&$-5.00_{-4.31}^{+3.98}$&Uniform&$-10^6$&$10^6$\\
$J^{\rm LC}$&m\,s$^{-1}$&RV jitter for LC&$4.71_{-2.07}^{+1.74}$&Uniform&0&$10^6$\\
$\gamma^{\rm UVES}$&m\,s$^{-1}$&RV offset for UVES&$-17.88_{-3.36}^{+2.96}$&Uniform&$-10^6$&$10^6$\\
$J^{\rm UVES}$&m\,s$^{-1}$&RV jitter for UVES&$0.61_{-0.09}^{+0.09}$&Uniform&0&$10^6$\\
$\gamma^{\rm VLC}$&m\,s$^{-1}$&RV offset for VLC&$-14.95_{-3.58}^{+3.13}$&Uniform&$-10^6$&$10^6$\\
$J^{\rm VLC}$&m\,s$^{-1}$&RV jitter for VLC&$0.94_{-0.60}^{+0.92}$&Uniform&0&$10^6$\\
ln$J_{\rm hip}$&---&Logarithmic jitter for hip&$0.20_{-0.13}^{+0.13}$&Uniform&-12&12\\
ln$J_{\rm gaia}$&---&Logarithmic jitter for gaia&$2.82_{-0.69}^{+0.79}$&Uniform&-12&12\\
\hline
\multicolumn{7}{p{1\textwidth}}{\footnotesize$^a$ The MA model of
  order $q$ or MA(q) is parameterized by the amplitudes of $q$ components
  and the logarithmic time scale ln$\tau$, where $\tau$ is in units of days.
  The MA model is introduced by \cite{tuomi12} and is frequently used in
  red-noise modeling for RV analyses (e.g., \citealt{tuomi13} and \citealt{feng16}). The superscripts of MA parameters and offsets represent the names of data sets. The definition of Gaia and Hipparcos jitter (lnJ$_{\rm gaia}$ and lnJ$_{\rm hip}$) can be found in \cite{feng19b}.}
\end{tabular}
\end{center}
\end{table}

\begin{table}
\caption{Other parameters for \eri b.}
\label{tab:other_HD22049}
\begin{center}
\begin{tabular}{lp{.10\textwidth}llllr}
\hline\hline
Parameter &Unit & Meaning& \eri b & Prior & Minimum & Maximum\\
\hline
$\gamma^{\rm APFp}$&m\,s$^{-1}$&RV offset for APFp&$0.68_{-0.37}^{+0.25}$&Uniform&$-10^6$&$10^6$\\
$J^{\rm APFp}$&m\,s$^{-1}$&RV jitter for APFp&$7.24_{-0.34}^{+0.35}$&Uniform&0&$10^6$\\
$\gamma^{\rm APFh}$&m\,s$^{-1}$&RV offset for APFh&$5.74_{-0.31}^{+0.34}$&Uniform&$-10^6$&$10^6$\\
$J^{\rm APFh}$&m\,s$^{-1}$&RV jitter for APFh&$4.93_{-0.27}^{+0.28}$&Uniform&0&$10^6$\\
$w_{1}^{\rm APFh}$&---&Amplitude of component 1 of MA(-Inf) for APFh&$0.87_{-0.06}^{+0.05}$&Uniform&-1&1\\
ln$\tau^{\rm APFh}$&---&Logarithmic time scale of MA(-Inf) for APFh&$0.06_{-0.33}^{+0.24}$&Uniform&-12&12\\
$\gamma^{\rm EXPRES}$&m\,s$^{-1}$&RV offset for EXPRES&$-6.96_{-0.21}^{+0.24}$&Uniform&$-10^6$&$10^6$\\
$J^{\rm EXPRES}$&m\,s$^{-1}$&RV jitter for EXPRES&$2.87_{-0.14}^{+0.15}$&Uniform&0&$10^6$\\
$w_{1}^{\rm EXPRES}$&---&Amplitude of component 1 of MA(2) for EXPRES&$0.87_{-0.07}^{+0.07}$&Uniform&-1&1\\
$w_{2}^{\rm EXPRES}$&---&Amplitude of component 2 of MA(2) for EXPRES&$0.14_{-0.08}^{+0.08}$&Uniform&NA&NA\\
ln$\tau^{\rm EXPRES}$&---&Logarithmic time scale of MA(2) for EXPRES&$1.19_{-0.20}^{+0.23}$&Uniform&-12&12\\
$\gamma^{\rm HARPN}$&m\,s$^{-1}$&RV offset for HARPN&$7.83_{-0.27}^{+0.37}$&Uniform&$-10^6$&$10^6$\\
$J^{\rm HARPN}$&m\,s$^{-1}$&RV jitter for HARPN&$13.57_{-0.55}^{+0.51}$&Uniform&0&$10^6$\\
$\gamma^{\rm HARPSpost}$&m\,s$^{-1}$&RV offset for HARPSpost&$-2.56_{-0.30}^{+0.21}$&Uniform&$-10^6$&$10^6$\\
$J^{\rm HARPSpost}$&m\,s$^{-1}$&RV jitter for HARPSpost&$4.87_{-0.28}^{+0.30}$&Uniform&0&$10^6$\\
$\gamma^{\rm HARPSpre}$&m\,s$^{-1}$&RV offset for HARPSpre&$-12.08_{-0.37}^{+0.36}$&Uniform&$-10^6$&$10^6$\\
$J^{\rm HARPSpre}$&m\,s$^{-1}$&RV jitter for HARPSpre&$1.47_{-0.05}^{+0.05}$&Uniform&0&$10^6$\\
$w_{1}^{\rm HARPSpre}$&---&Amplitude of component 1 of MA(3) for HARPSpre&$0.78_{-0.04}^{+0.04}$&Uniform&-1&1\\
$w_{2}^{\rm HARPSpre}$&---&Amplitude of component 2 of MA(3) for HARPSpre&$0.17_{-0.05}^{+0.05}$&Uniform&-1&1\\
$w_{3}^{\rm HARPSpre}$&---&Amplitude of component 3 of MA(3) for HARPSpre&$0.05_{-0.04}^{+0.04}$&Uniform&-1&1\\
ln$\tau^{\rm HARPSpre}$&---&Logarithmic time scale of MA(3) for HARPSpre&$0.10_{-0.20}^{+0.20}$&Uniform&-12&12\\
$\gamma^{\rm KECK}$&m\,s$^{-1}$&RV offset for KECK&$5.05_{-0.31}^{+0.34}$&Uniform&$-10^6$&$10^6$\\
$J^{\rm KECK}$&m\,s$^{-1}$&RV jitter for KECK&$6.86_{-0.21}^{+0.22}$&Uniform&0&$10^6$\\
$w_{1}^{\rm KECK}$&---&Amplitude of component 1 of MA(1) for KECK&$0.95_{-0.07}^{+0.03}$&Uniform&-1&1\\
ln$\tau^{\rm KECK}$&---&Logarithmic time scale of MA(1) for KECK&$1.56_{-0.19}^{+0.20}$&Uniform&-12&12\\
$\gamma^{\rm KECKj}$&m\,s$^{-1}$&RV offset for KECKj&$-0.22_{-0.19}^{+0.20}$&Uniform&$-10^6$&$10^6$\\
$J^{\rm KECKj}$&m\,s$^{-1}$&RV jitter for KECKj&$6.43_{-0.33}^{+0.37}$&Uniform&0&$10^6$\\
$\gamma^{\rm LC}$&m\,s$^{-1}$&RV offset for LC&$5.17_{-0.55}^{+0.43}$&Uniform&$-10^6$&$10^6$\\
$J^{\rm LC}$&m\,s$^{-1}$&RV jitter for LC&$5.27_{-0.33}^{+0.33}$&Uniform&0&$10^6$\\
$w_{1}^{\rm LC}$&---&Amplitude of component 1 of MA(1) for LC&$0.75_{-0.11}^{+0.11}$&Uniform&-1&1\\
ln$\tau^{\rm LC}$&---&Logarithmic time scale of MA(1) for LC&$2.65_{-0.27}^{+0.29}$&Uniform&-12&12\\
$\gamma^{\rm LICK13}$&m\,s$^{-1}$&RV offset for LICK13&$5.67_{-0.42}^{+0.56}$&Uniform&$-10^6$&$10^6$\\
$J^{\rm LICK13}$&m\,s$^{-1}$&RV jitter for LICK13&$4.88_{-0.34}^{+0.48}$&Uniform&0&$10^6$\\
$\gamma^{\rm LICK6}$&m\,s$^{-1}$&RV offset for LICK6&$0.38_{-0.22}^{+0.19}$&Uniform&$-10^6$&$10^6$\\
$J^{\rm LICK6}$&m\,s$^{-1}$&RV jitter for LICK6&$8.14_{-0.38}^{+0.60}$&Uniform&0&$10^6$\\
$\gamma^{\rm LICK8}$&m\,s$^{-1}$&RV offset for LICK8&$-0.42_{-0.13}^{+0.10}$&Uniform&$-10^6$&$10^6$\\
$J^{\rm LICK8}$&m\,s$^{-1}$&RV jitter for LICK8&$9.22_{-0.28}^{+0.28}$&Uniform&0&$10^6$\\
$w_{1}^{\rm LICK8}$&---&Amplitude of component 1 of MA(2) for LICK8&$0.62_{-0.10}^{+0.11}$&Uniform&-1&1\\
$w_{2}^{\rm LICK8}$&---&Amplitude of component 2 of MA(2) for LICK8&$0.30_{-0.13}^{+0.13}$&Uniform&-1&1\\
ln$\tau^{\rm LICK8}$&---&Logarithmic time scale of MA(2) for LICK8&$2.48_{-0.33}^{+0.31}$&Uniform&-12&12\\
$\gamma^{\rm VLC}$&m\,s$^{-1}$&RV offset for VLC&$3.15_{-0.37}^{+0.33}$&Uniform&$-10^6$&$10^6$\\
$J^{\rm VLC}$&m\,s$^{-1}$&RV jitter for VLC&$1.41_{-0.55}^{+0.40}$&Uniform&0&$10^6$\\
$w_{1}^{\rm VLC}$&---&Amplitude of component 1 of MA(1) for VLC&$0.80_{-0.12}^{+0.10}$&Uniform&-1&1\\
ln$\tau^{\rm VLC}$&---&Logarithmic time scale of MA(1) for VLC&$3.98_{-0.43}^{+0.31}$&Uniform&-12&12\\
ln$J_{\rm hip}$&---&Logarithmic jitter for hip&$-5.61_{-0.40}^{+0.38}$&Uniform&-12&12\\
ln$J_{\rm gaia}$&---&Logarithmic jitter for gaia&$2.00_{-0.38}^{+0.36}$&Uniform&-12&12\\
\hline
\end{tabular}
\end{center}
\end{table}

\section{New APF RVs for \eri}

\begin{table}
 \caption{APF data for \eri}
 \label{tab:apfp}
    \begin{tabular}{lccc}
      \textbf{BJD} & \textbf{RV [m/s]} & \textbf{RV error [m/s]} & \textbf{S-index}\\
      \hline
       2456582.93034 & 26.64 & 2.73 & 0.524\\
      2456597.91368 & 6.40 & 2.36 & 0.528\\
      2456606.68427 & 16.52 & 0.75 & 0.531\\
      2456608.10376 & 4.69 & 0.78 & 0.530\\
      2456610.76250 & 16.04 & 1.18 & 0.512\\
      2456618.88476 & -2.11 & 0.78 & 0.530\\
      2456624.72004 & 4.20 & 1.11 & 0.519\\
      2456626.81421 & 24.46 & 0.75 & 0.521\\
      2456628.72976 & 24.14 & 0.70 & 0.540\\
      2456631.42746 & -2.26 & 0.88 & 0.502\\
      2456632.80921 & 14.46 & 0.62 & 0.523\\
      2456644.75696 & 8.20 & 2.30 & 0.522\\
      2456647.81171 & 14.44 & 0.63 & 0.535\\
      2456648.59184 & 12.62 & 1.10 & 0.538\\
      2456662.63738 & 9.77 & 0.73 & 0.536\\
      2456663.75415 & 10.43 & 1.11 & 0.531\\
      2456667.52792 & 18.00 & 0.78 & 0.535\\
      2456671.68695 & 19.96 & 1.05 & 0.604\\
      2456675.75647 & 7.84 & 1.12 & 0.519\\
      2456679.83732 & 17.70 & 1.05 & 0.529\\
      2456682.56608 & 17.80 & 0.82 & 0.550\\
      2456689.76638 & 26.34 & 0.75 & 0.500\\
      2456875.02028 & 7.12 & 2.18 & 0.501\\
      2456894.88054 & 8.28 & 1.30 & 0.470\\
      2456901.06193 & 9.95 & 1.54 & 0.479\\
      2456909.10279 & -4.71 & 1.21 & 0.476\\
      2456922.07953 & 12.25 & 2.13 & 0.461\\
      2456935.94021 & -2.43 & 1.27 & 0.479\\
      2456937.92403 & -0.55 & 1.35 & 0.468\\
      2456950.03798 & 3.82 & 1.44 & 0.472\\
      2456985.64755 & -1.80 & 2.28 & 0.441\\
      2456988.63095 & 5.93 & 1.29 & 0.478\\
      2456999.76434 & 8.84 & 1.37 & 0.459\\
 2457015.72916 & -2.17 & 1.10 & 0.465 \\ 
 2457026.78021 & -1.44 & 1.34 & 0.464 \\
 2457058.45996 & -3.69 & 1.89 & 0.435 \\
 2457234.08236 & 7.73 & 1.39 & 0.525 \\
 2457245.86234 & -4.19 & 1.41 & 0.519 \\
 2457249.93007 & -3.94 & 1.31 & 0.500 \\
 2457253.11257 & 5.63 & 1.33 & 0.511 \\
 2457257.15719 & -1.02 & 1.15 & 0.506 \\
 2457258.94437 & -12.69 & 1.23 & 0.517 \\
 2457261.02221 & -2.76 & 1.32 & 0.501 \\
 2457262.94505 & -7.81 & 1.36 & 0.496 \\
 2457265.95783 & 9.67 & 1.24 & 0.516 \\
 2457275.01304 & -1.91 & 1.23 & 0.515 \\
 2457283.96368 & 1.88 & 1.29 & 0.507 \\
 2457287.02735 & -1.11 & 1.35 & 0.524 \\
 2457290.95635 & 3.19 & 1.42 & 0.534 \\
 2457305.83659 & -5.63 & 1.23 & 0.515 \\
 2457308.90844 & 13.30 & 1.26 & 0.534 \\
 2457318.83435 & 8.72 & 1.26 & 0.557 \\
 2457321.79157 & 6.64 & 1.36 & 0.540 \\
 2457325.84352 & 2.87 & 1.41 & 0.543 \\
 2457331.10764 & 9.90 & 1.36 & 0.552 \\
 2457332.78237 & 9.64 & 1.25 & 0.558 \\
 2457334.82998 & 5.22 & 1.30 & 0.548 \\
 2457337.78910 & 5.41 & 1.59 & 0.545 \\
 2457340.95644 & -1.99 & 1.27 & 0.553 \\
 2457347.86896 & 4.10 & 1.29 & 0.556 \\
 2457348.77993 & 4.65 & 1.27 & 0.556 \\
 2457350.72611 & 5.83 & 1.20 & 0.558 \\
 2457354.70613 & -0.88 & 1.65 & 0.548 \\
 2457361.64656 & 17.26 & 1.43 & 0.549 \\
 \hline
\end{tabular}
\end{table}

\bsp	
\label{lastpage}
\end{document}